\documentclass[12pt]{iopart}
\input{iopams.sty}

\usepackage{graphicx}
\usepackage{hyperref}

\begin{document}

\title{High-pressure behaviour of GeO$_2$: a simulation study.}

\author {Dario Marrocchelli$^1$, Mathieu Salanne$^{2,3}$, Paul A Madden$^4$}

\address{$^1$ School of Chemistry, University of Edinburgh, Edinburgh EH9 3JJ, United Kingdom}
\address{$^2$ UPMC Univ Paris 06, UMR 7612, PECSA, F-75005 Paris, France}
\address{$^3$ CNRS, UMR 7612, PECSA, F-75005 Paris, France}
\address{$^4$ Department of Materials, University of Oxford, Parks Road, Oxford OX1 3PH, United Kingdom}

\begin{abstract}
In this work we study the high pressure behaviour of liquid and glassy GeO$_2$ by means of molecular dynamics simulations. The interaction potential, which includes dipole polarization effects, was parameterized from first-principles calculations. Our simulations reproduce the most recent experimental structural data very well. The character of the pressure induced structural transition in the glassy system has been a matter of controversy. We show that our simulations and the experimental data are consistent with
a smooth transition from a tetrahedral to octahedral network with a significant number of penta-coordinated germanium ions appearing over an extended pressure range. 
Finally, the study of high-pressure, liquid germania confirms that this material presents an anomalous behaviour of the diffusivity as observed in analogous systems such as silica and water. The importance of penta-coordinated germanium ions for such behaviour is stressed.
\end{abstract}

\maketitle

\section{Introduction}
Germania (GeO$_2$), along with silica and  beryllium fluoride, is a ``strong"
glass-former \cite{angell1995a} characterized by a tetrahedral network structure at
ambient conditions in the amorphous phase~\cite{galeener1983a}. Its structure is
based on corner-sharing Ge(O$_{1/2}$)$_4$ tetrahedra (the $1/2$ index means that
each O$^{2-}$ ion is shared by two Ge$^{4+}$), with a Ge-O average distance of
1.73~\AA\ and a mean inter-tetrahedral Ge-O-Ge angle of
132$^\circ$~\cite{micoulaut2006b,salmon2007a}. There is considerable interest in
the behaviour of such tetrahedrally coordinated glasses under pressure. In the associated crystalline materials
high-pressure transitions are observed between four- and six-coordinate structures.
Pressure-induced structural changes in the amorphous phases have been linked to
anomalous behaviour in the elastic, viscous and thermal properties and to the phenomenon
of polyamorphism \cite{mcmillan2009a}. In amorphous silica itself, the major structural
transition occurs at relatively high pressures ($\sim$25 GPa) where direct
structural studies remain difficult \cite{wilding2008a}. However, because of the larger
cation / anion radius ratio in germania relative to silica, the transition occurs in
a pressure domain now accessible to structural studies. Despite this, a clear
picture of the nature of the pressure-induced structural changes in germania has only recently started to emerge. We have therefore performed
computer simulations in an attempt to clarify the relationship of the information
emerging from the different experiments.

High-pressure structural studies have been made using EXAFS
spectroscopy~\cite{itie1989a,vaccari2009a, hong2009a, baldini2010a} and X-ray and neutron diffraction
experiments~\cite{salmon2007a, guthrie2004a, drewitt2010a}. All the experimental studies agreed on the
existence of a structural change in the pressure range 3-15 GPa, associated with an
increase in the Ge-O separation which is broadly consistent with an increase in coordination number. Vaccari {\it et al.} \cite{vaccari2009a}, for example, showed, from EXAFS studies, that this distance
switched from 1.74~\AA\ (at 0~GPa) to 1.82~\AA\ (at 13~GPa). On the other hand, the
literature has long been very contradictory about the character of the change in coordination
number. The first study to address this issue is that of Itie {\it et al.} \cite{itie1989a}
who performed x-ray absorption measurements up to 23.2 GPa and reported that the Ge coordination
changes from fourfold to sixfold at pressure between 7 and 9 GPa.
 Vaccari {\it et al.}, on the other hand, proposed a progressive shift in the coordination number,
and found no evidence of a fully six-coordinate structure, even at the highest pressure of their study (13 $\sim$ GPa).
Very recent XAFS and EXAFS studies \cite{hong2009a, baldini2010a} extended the studied pressure range to 53 and 44 GPa respectively and it was postulated that a complete 6-fold coordination of the Ge ions is only reached at pressures as high as 25-30 GPa. The ability of EXAFS to provide very accurate first-neighbour distances is
well established, but so is its limitation for the determination of coordination
numbers in amorphous materials~\cite{filipponi1995a,filipponi2001a,okamoto2004a}. Only an average coordination
number can be extracted, with an important error bar which depends crucially on
assumptions about the shape of the radial distribution functions at larger
separations, and to estimate proportions of different coordination polyhedra in a material from
EXAFS alone is not normally considered reliable.

In principle neutron diffraction, when employing the isotopic substitution method, gives access to all the partial
structure factors, and hence to the corresponding partial radial distribution
functions (RDF) which contain all the structural information. In the case of
GeO$_2$, this programme has been fulfilled only at ambient pressure by Salmon {\it
et al.}~\cite{salmon2006a}: at elevated pressures only total X-ray and neutron diffraction patterns are available.
A first attempt to study the high pressure system by these means was reported by Guthrie {\it et al} \cite{guthrie2004a}: unfortunately, the data were too noisy to
extract good RDFs. However, a very sharp transition from a tetrahedral to an octahedral structure was proposed,
with a small range of pressure in which the system is characterized by five-fold
coordination around Ge$^{4+}$ ions.  Very recently, Drewitt {\it et al.} have performed new neutron diffraction
measurements up to 8.6 GPa and obtained data of very high quality \cite{drewitt2010a}. Their results show
a gradual change of the intermediate range order with increasing density as manifested
by an increase in position and reduction in height of the first sharp diffraction peak in the
total structure factor. From their data there is no evidence in support of an abrupt
transformation of the network structure over the investigated pressure range, in agreement with
the most recent EXAFS experiments \cite{vaccari2009a, hong2009a, drewitt2010a}.

Computer simulations could help in resolving  these differences since they can
provide a detailed picture of the structure of a system. Unfortunately, to date,
all the molecular dynamics (MD) calculations involving pressurized germania have
been performed with pair potentials of limited accuracy
~\cite{micoulaut2004b,gutierrez2004a,hoang2006a,shanavas2006a,hung2007a,hoang2007a, peralta2008a, hung2008a, li2009b}.
For example, for the most commonly used potential, developed by Oeffner and Elliot~\cite{oeffner1998a},
{\em two} different parameter sets were proposed: an original one, which was fitted from an ab initio energy surface, and a so-called rescaled one, which was developed from the previous one in order to give a better reproduction of the
vibrational properties.  For this reason, some apparent inconsistencies have arisen in the literature because different classical potentials were being used \cite{micoulaut2004b, peralta2008a, hawlitzky2008a,marrocchelli2009a}. Finally, since the OE potential is not able to reproduce with a single set of parameters all the ambient pressure properties of amorphous
germania, it is also reasonable to suspect that it might fail when studying the pressure behaviour of this material. This might explain the disagreement with the experimental trends found in some papers in the literature. For instance, in ref. \cite{li2009b}, the Ge-O bond distance is found to remain constant in the 3-25 GPa pressure
interval, whilst in ref. \cite{shanavas2006a} the Ge-O coordination number is found to increase continuously from the lowest pressure (1GPa), both trends being in strong contrast with the experimental evidence \cite{itie1989a, vaccari2009a, guthrie2004a}.

Recently, first-principles molecular dynamics (FPMD) studies of glassy GeO$_2$ have also been reported~\cite{peralta2008a, giacomazzi2005a, giacomazzi2006a, zhu2009a}. In principle, the amount of empirical information needed to set up a first-principles calculation is minimal and it would normally be the method of choice to study the physico-chemical properties of condensed phase systems. However, it is computationally very expensive compared with classical molecular dynamics, which is a major drawback when dealing with glassy systems, where the structural relaxation times are necessarily long. These studies have therefore focused on the study of the structural and vibrational properties of the glass; only one of them included results for high pressure systems~\cite{zhu2009a}. \newline

We recently introduced a new interaction
potential, which includes many-body polarization effects, and succeeded in
reproducing the structural {\em and} vibrational properties of glassy germania at ambient pressure as well as
the dynamical properties of liquid germania in the 3600-5000 K temperature range~\cite{marrocchelli2009a}, with a {\em single} set of parameters.
This potential was parameterized with reference to extensive Density Functional Theory (DFT)
electronic structure calculations  on disordered condensed-phase configurations. Also, similarly constructed potentials have been shown to provide an excellent, transferable description of a number of complex oxides \cite{norberg2009a,jahn2007b}. The aim of this paper is to show how the use of this potential together with a direct comparison
with the available experimental data, can improve our understanding
of the behavior of glassy germania at high pressure. In the end, we will emphasize
that the various experimental studies are in accord, even if the
initial interpretations that were made of them did not.

Once the question of structure is resolved,  it is of interest to study the
influence of structural changes on the physical properties of the system. A
well-known effect is the existence of dynamical anomaly in silica-based systems
when the pressure increases \cite{mcmillan2009a}. Unlike most other systems which, under
compression, tend to show a decreasing mobility of the species, in liquid silica
the diffusion coefficients increase until they reach a maximum for a given
pressure, and then decrease. This anomaly was linked to the existence of
pentacoordinated species~\cite{angell1982a}. The only evidence for such a behaviour in
liquid germania was provided by Sharma {\it et al.} who showed a decrease of the
viscosity with pressure, though the pressure range of this study was very limited
(0-1~GPa)~\cite{sharma1979a}. There are also some simulations studies on high-pressure liquid germania
but, unfortunately, the results are contradictory. Hoang {\it et al.} \cite{hoang2006a} do indeed observe an anomalous
behaviour of the diffusivity in GeO$_2$. However, Hung and co-workers \cite{hung2007a}, who also studied this system
by means of classical molecular dynamics, found no evidence of such a phenomenon, but again both of these studies involved classical pairwise additive interaction potentials of limited accuracy.
To address this issue we have therefore studied the structural and
dynamical properties of liquid germania, at a temperature of 4000~K and over a wide range of pressures.

\section{Simulation details}
The interaction potential used in this  study has already been described
elsewhere~\cite{marrocchelli2009a}. The ionic species carry their valence
charges (Ge$^{4+}$ and O$^{2-}$), and the polarization effects that result from the
induction of dipoles on the oxide ions are accounted for. The parameters of the
interaction potential for germania were obtained from the application of a force-
and dipole-matching procedure aimed at reproducing a large set of first-principles
(DFT) reference data \cite{madden2006a} on the condensed phase. 

 To obtain structural properties in the glassy state, we performed MD
simulations in the $NPT$ ensemble using the method introduced by Martyna {\it et
al.}~\cite{martyna1994a}. Details
on the generation of a compressed glassy state will be reported below.
 When dynamical properties were computed, the system was
simulated in the $NVT$ ensemble using the Nos\'e-Hoover chain thermostat
method~\cite{martyna1992a}. A relaxation time of 10~ps was used for the thermostats
and the barostats. The simulations on glassy GeO$_2$ ($T$~=~300~K) were performed on a
simulation cell containing 432 atoms whereas the NVT simulation cells on liquid
germania ($T$~=~4000~K) contained 600 atoms. We used a time step of 1~fs to integrate the equations of motion, and
minimization of the polarization energy was carried out at each time step using a
conjugate gradient method. All simulations were equilibrated for at least 0.5~ns and
then a subsequent run of 0.5~ns was made to accumulate enough statistics. 

\section{Compressing a glass: some limitations}
Generating a glassy state by means of computer simulations is a challenging operation. Indeed the relatively short time-scales available
(a few ns in our case) force us to use unrealistically fast cooling rates. This implies that sometimes the glass we obtained is further away from equilibrium
than the experimental one. In fact, even experimentally, glasses prepared in different ways show different properties, implying that even on a time-scale of hours ({\it i.e.} 12 orders of magnitude higher than what we can afford with MD simulations) there are still some relaxation effects.
This problem becomes even more significant when one tries to compress a glass. To this end, Scandolo {\it et al.} described two different compression methodologies for SiO$_2$~\cite{liang2008a}. The first one is a cold compression route consisting of increasing pressure slowly at ambient temperature while the second one, which they called a `quench-from-the-melt' procedure, consists of obtaining the compressed glass by a slow
cooling of a compressed sample from a temperature where atomic diffusion is observable on the time scale of the MD simulation. By comparing their results with the available experimental
data on compressed glass, they showed that the samples obtained with the two procedures are representative of the experimental {\it in situ} compressed
(as, for example, in refs. \cite{vaccari2009a, hong2009a}) and densified forms of glass \cite{stone2001a}, respectively. They found that
structural differences between annealed and cold-compressed forms were most noticeable at distances of 3.5--4 \AA. \newline

\begin{figure}
\resizebox{\textwidth}{!}{\includegraphics{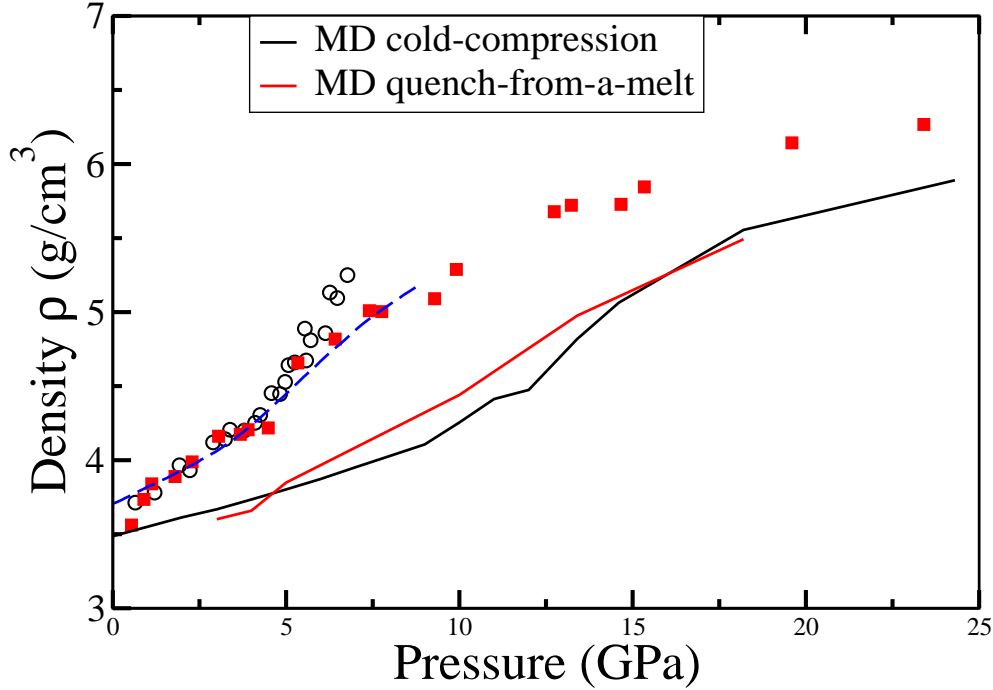}}
\caption{ The mass density $\rho$ for GeO$_2$ glass as a function of pressure for cold compression and quench-from-a-melt procedures. Experimental data are from the {\it in-situ} compression studies of Hong et al. \cite{hong2007a} ((red squares)), Smith et al. \cite{smith1995a} (empty circles) and Tsiok et al. \cite{tsiok1998a} (blue broken curve).}
\label{figure1}
\end{figure}

In the case of GeO$_2$ there has been little effort so far in this direction. Most of the previous work on this system used the cold compression route but a direct quantitative comparison with experimental data was never attempted. In this work we tried both routes used by Scandolo {\it et al.} and compared the results with the experimental density vs pressure data. This is shown in figure \ref{figure1}. It can be appreciated that neither of the two routes reproduces quantitatively the experimental data. In both cases the simulated glass seems to be less responsive to compression, which we attribute to the limited simulation times (see below). It seems, however, that the quench-from-a-melt procedure gives a closer agreement with the experimental data than the cold-compressed route, once again indicating that a certain degree of diffusion helps the glass relaxation. It has to be remembered, however, that Scandolo {\it et al.} showed how the quench-from-a-melt route is more representative of the densified forms of the glass, so that a comparison with the experimental data on {\it in situ} compressed glass, such as in refs. \cite{guthrie2004a, salmon2006a} might not work as well. \newline

It should be evident from the above discussion that we are unable to reproduce quantitatively the equation of state of glassy germania. This should not necessarily be taken as a limitation of our potential but as a consequence of the relatively short time-scales available in computer simulations. For this reason, in the remainder of this paper, we will simulate cold-compressed GeO$_2$ and report our data as a function of {\it density}, instead of pressure. When a comparison with experimental data is required, we will convert the experimental pressures into densities by using the data in  figure \ref{figure1} from ref. \cite{tsiok1998a,hong2007a}. In the case of liquid germania, since the the system is very diffusive and reaches equilibrium in a few picoseonds, we do not anticipate any problem with the equation of state and we will therefore report our data as a function of pressure.

\begin{figure}
\resizebox{\textwidth}{!}{\includegraphics{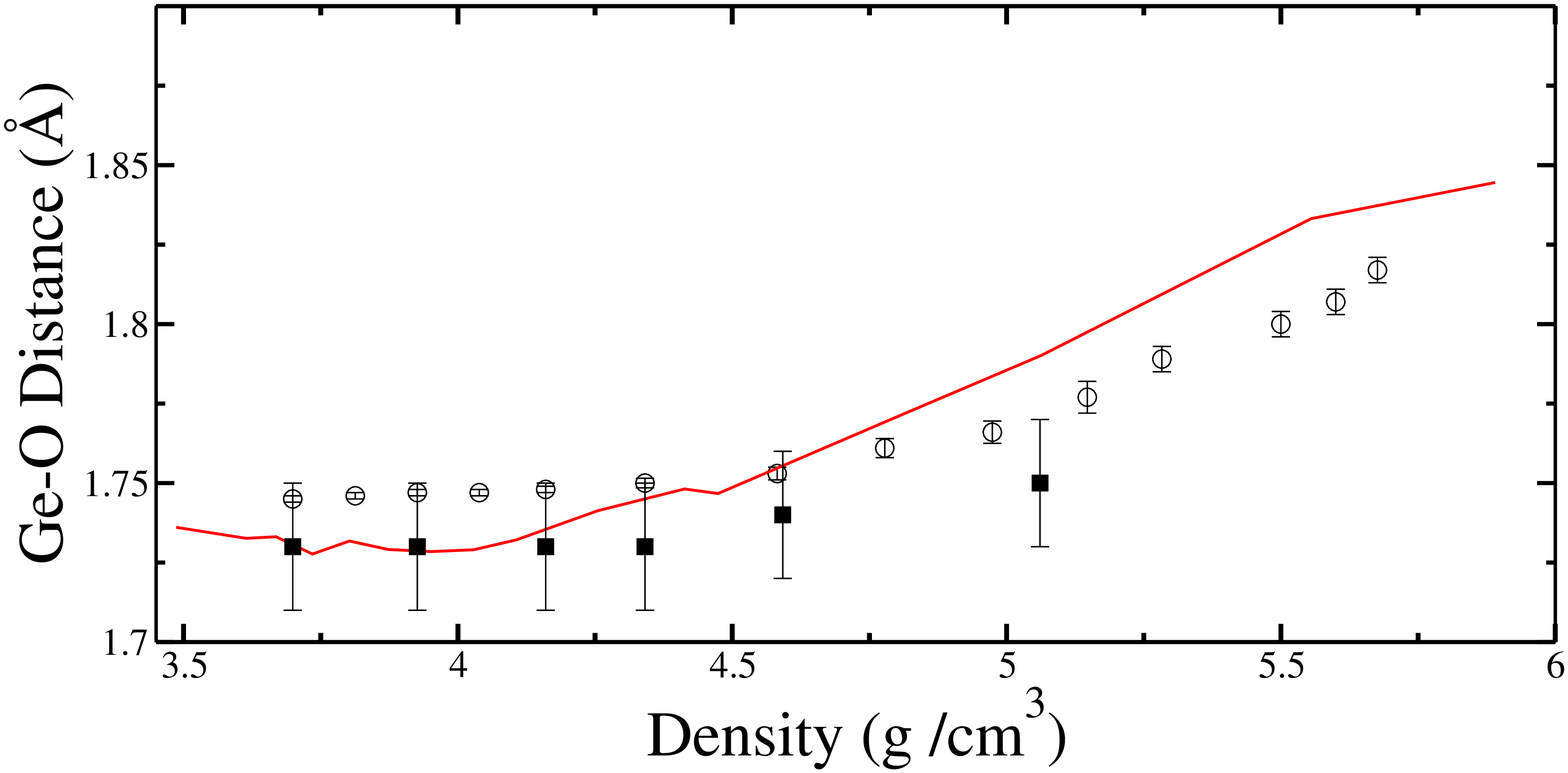}}
\resizebox{\textwidth}{!}{\includegraphics{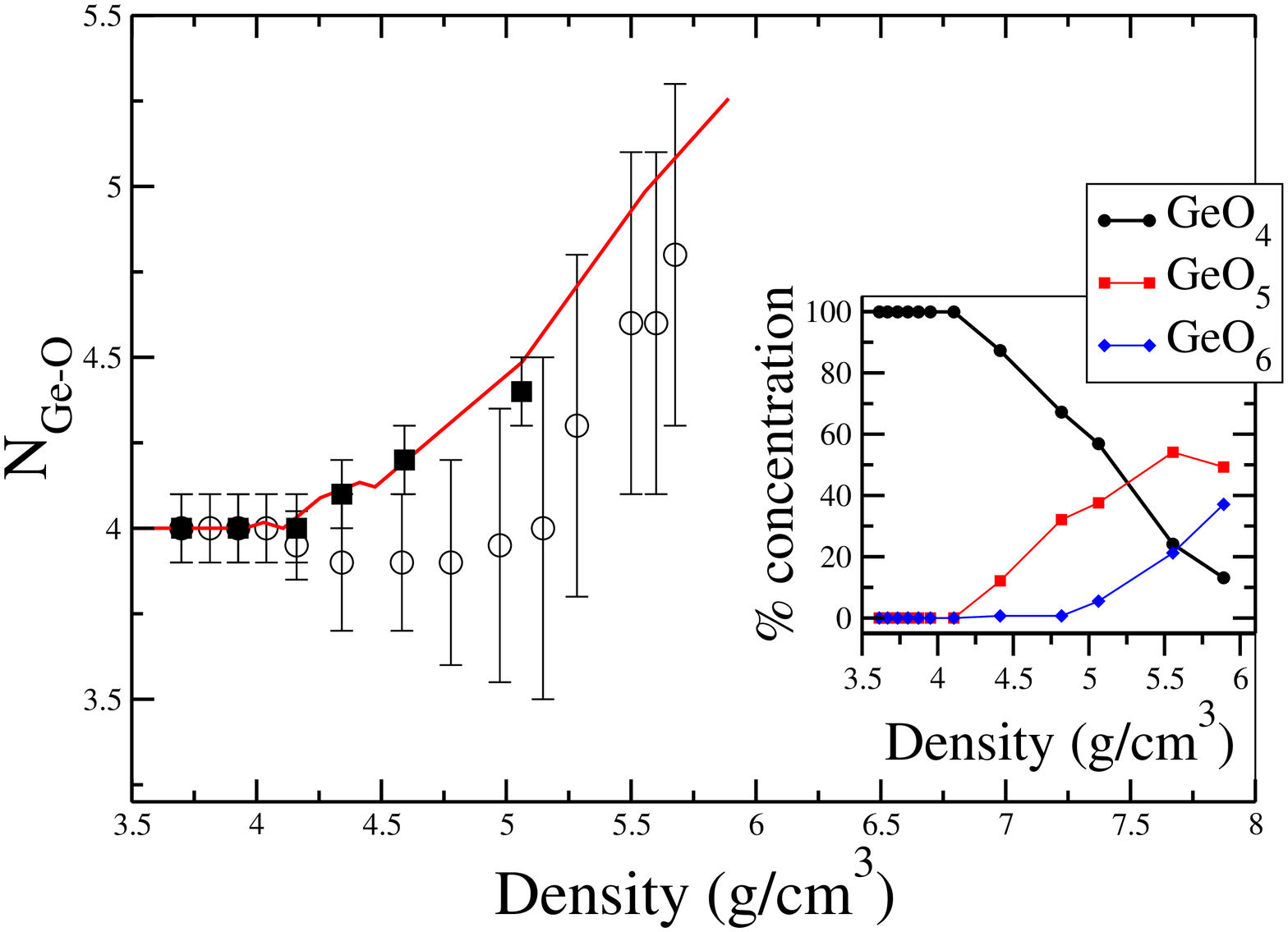}}
\caption{ Top panel: comparison between the simulated (solid line) and experimental
Ge-O bond distance from Vaccari \cite{vaccari2009a}  (black empty dots) and Drewitt \cite{drewitt2010a} (black filled squares)
as a function of increasing density.
Bottom panel: comparison  between the simulated (solid line) and
experimental, (black empty dots) Vaccari \cite{vaccari2009a}  and  (black filled squares) Drewitt \cite{drewitt2010a}, first shell Ge-O coordination number.
The inset in the bottom panel shows the density dependence of the
percentage concentration of GeO$_4$, GeO$_5$, GeO$_6$ units. } \label{figure2}
\end{figure}

\section{Glassy germania}
In the top panel of figure \ref{figure2} we report the average Ge-O distance\footnote{This distance was
defined as $d_{GeO}=\frac{\int_0^{r_{cut}} r^3 g(r)dr}{\int_0^{r_{cut}} r^2 gr(r) dr}$
where g(r) is the Ge-O radial distribution function and r$_{cut}$ is the position of the first
minimum in the g(r)} as a function of increasing density. It is
compared to the values obtained by Vaccari {\it et al.} \cite{vaccari2009a} and Drewitt {\it et al.} \cite{drewitt2010a} {\it via} EXAFS
and neutron diffraction measurements.  The agreement is quite good, considering the scatter of the experimental data
and the intrinsic error associated with the pressure-to-density conversion. The simulations, like the experiments, show first a  Ge-O bond distance which is fairly constant in the 3.66 - 4.5 g/cm$^3$ density range (corresponding to an experimental pressure range of  0-5 GPa) and then
a gradual increase in the Ge-O bond distance, starting at a density of 4.5 g/cm$^3$.

We can interpret that data in terms of compression mechanisms. When a
tetrahedral system like GeO$_2$ is compressed, three types of structural change
could occur. Firstly, the system could keep exactly the same structure, but
adapt by allowing the bond lengths to decrease: obviously   this is not the case
here. Secondly, the tetrahedral units could be kept unmodified, in which case the
decrease of accessible volume would imply a rearrangement of the network structure,
progressively minimizing the size of the voids~\cite{wilson1998b, stone2001a} and resulting in changes in the first-sharp diffraction peak. Finally the
tetrahedral structure could be lost, and higher-coordinated structural units could
be formed. Such an evolution may induce an increase of the Ge-O bond length in
order to allow more than 4 oxide ions into the germanium first coordination shell.
The analysis of bond length data gives evidence of a succession of the two latter
mechanisms, at low and high pressures. This was also confirmed by an analysis of the
bond angle distributions (not shown).

To verify this, we have determined the germanium coordination number. The cut-off
used for the coordination number calculation was 2.38~\AA\, which corresponds to
the first minimum of the Ge-O partial RDF. Again we compare these values with the
ones obtained by Vaccari {\it et al.} \cite{vaccari2009a} and Drewitt  {\it et al.} \cite{drewitt2010a}. Good agreement is found, especially
with the neutron data. The simulated coordination number remains constant, at a value
of 4, until a density of 4.5 g/cm$^3$. In this first r\'egime, germania is
therefore keeping its tetrahedral structure under compression, but the tetrahedra
reorient themselves. 
Then, in a second r\'egime, more highly coordinated germanium ions begin
to be observed. In the inset of the figure, the percentage of GeO$_4$, GeO$_5$ and
GeO$_6$ units are given as a function of density. Once again, at a density of 4.5 g/cm$^3$,
some Ge$^{4+}$ ions start to accept 5 O$^{2-}$ ions in their vicinity.
Then, at $\rho \sim$  5.0 g/cm$^3$ (corresponding to an experimental pressure of about 7-8 GPa), GeO$_6$ units begin to be formed. At the highest density
studied experimentally, $\rho \sim$  5.85 g/cm$^3$ (corresponding to approximately 15 GPa), a non-negligible proportion of GeO$_4$ units is still observed.
We find no evidence of a state in which there are only 5-fold coordinated Ge ions in the studied pressure range. These
fivefold-coordinated units play an important role in liquid germania as we shall show in the next section. Our results can also be compared to the recent FPMD simulations on the same system~\cite{zhu2009a}. In the latter, a much sharper change in the first-neighbour distances is observed upon compression: At a density of 5.4~g.cm$^{-3}$, the Ge-O distance already reaches a value of 1.89~\AA, showing important differences with experimental data. This is very likely due to the small size of the samples (108 atoms) and to the higher quenching rates that have to be involved in FPMD simulations.

\begin{figure}
\resizebox{\textwidth}{!}{\includegraphics{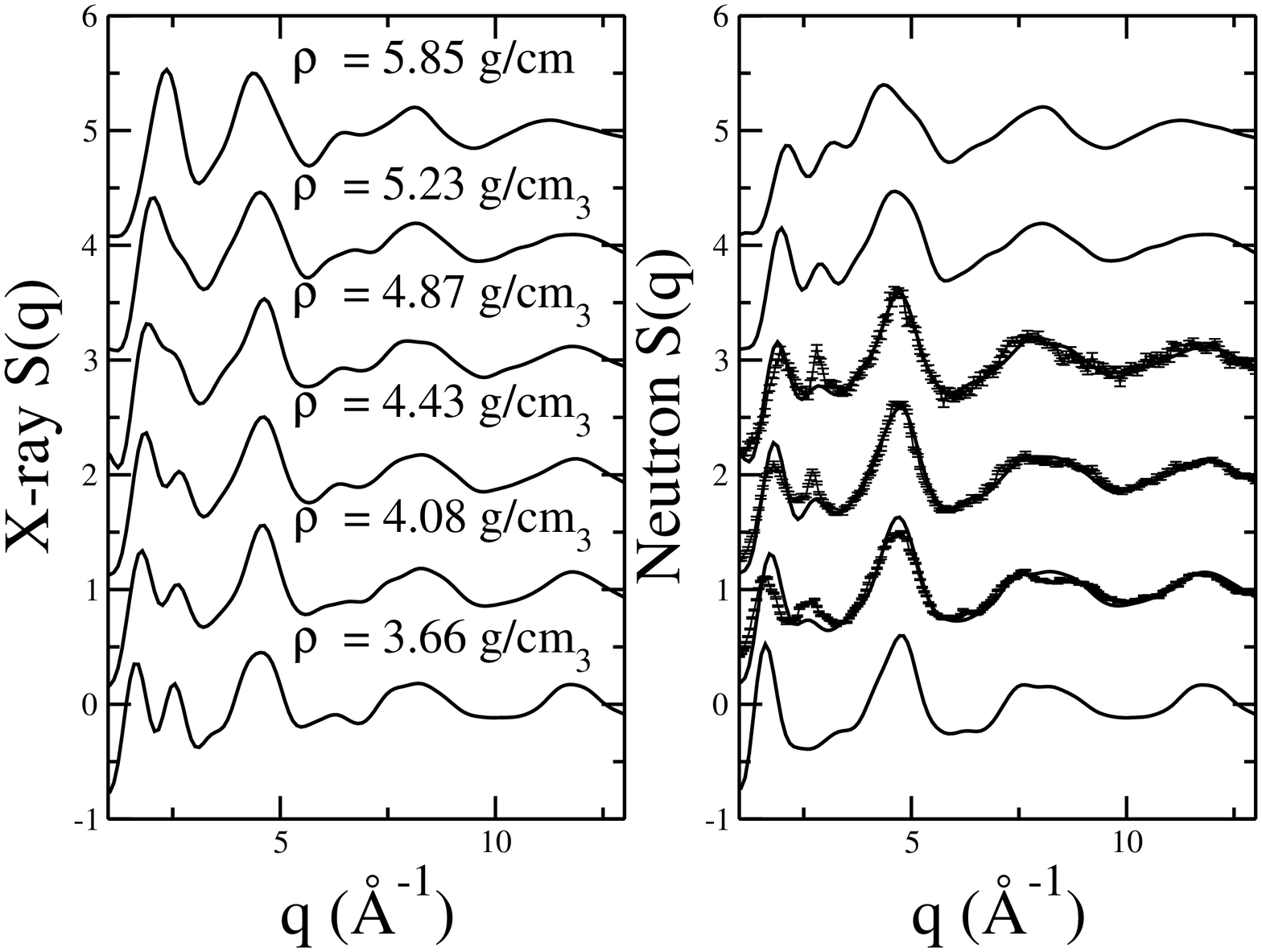}}
\resizebox{\textwidth}{!}{\includegraphics{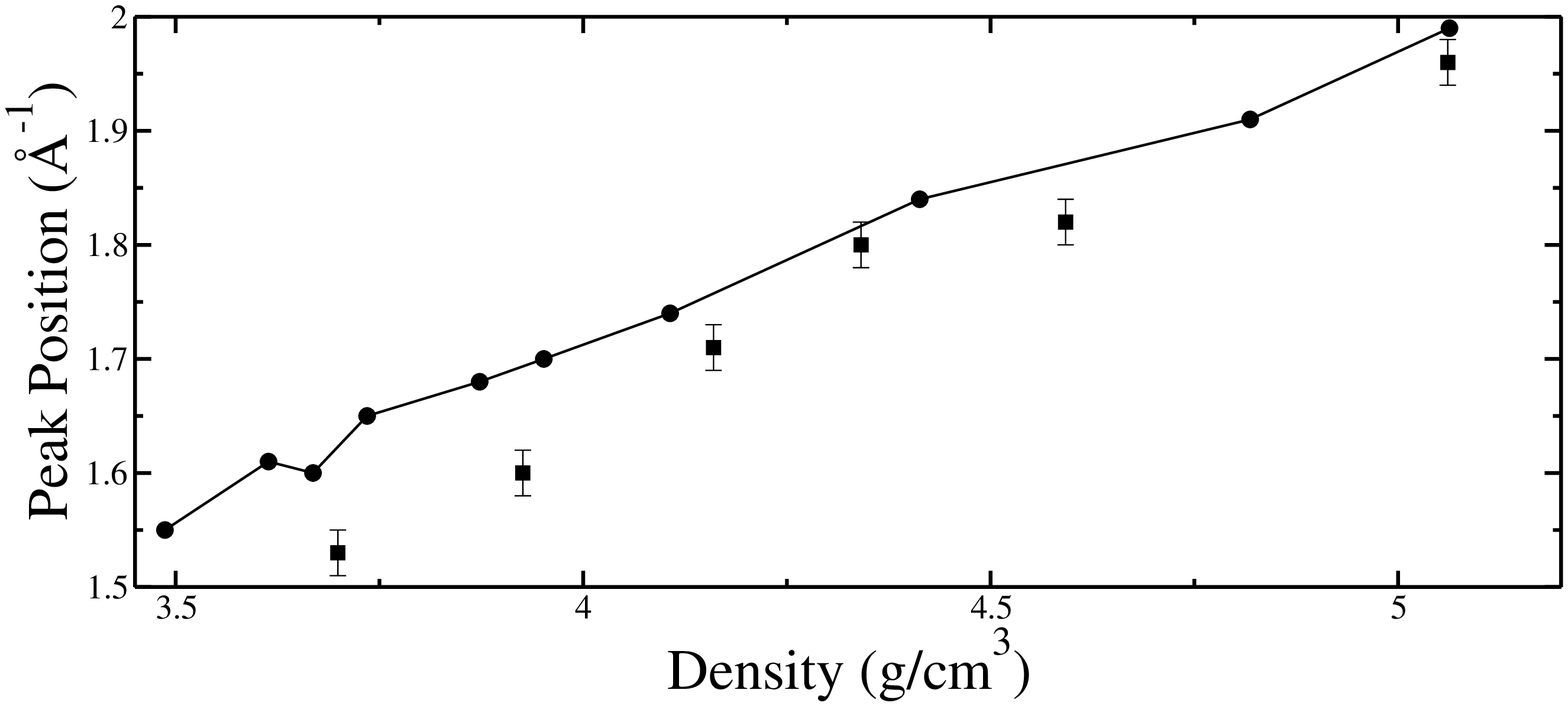}}
\caption{ Top: simulated X-ray and  neutron diffraction patterns as a function of
increasing density. The experimental data in the right-hand panel (line with error bar) are from \cite{drewitt2010a}.
Bottom: pressure dependence of the FSDP (extracted from the neutron patterns) position obtained from the X-ray diffaction pattern. The experimental data (filled green squares) are from \cite{drewitt2010a}.}
\label{figure3}
\end{figure}

At first sight, our results  differ significantly from the  conclusions of Guthrie {\it et al.} from
their combined X-ray and neutron diffraction study of glassy germania \cite{guthrie2004a} .
Qualitatively, they observed an analogous mechanism, but with a much sharper
increase in the coordination number. They proposed a complete transition to a
GeO$_5$ units based structure for pressures ranging from 6 to 10~GPa, and a final
completely octahedral structure at a pressure of 15~GPa. These values were
extracted from a Fourier transform of the X-ray diffraction {\em total} structure
factor (their neutron data being too noisy to allow an attempt to extract
partials). To show that the discrepancy observed is due to the analysis of the
X-ray data, rather than a deficiency of the representation of the structure in the
simulations, we have computed the X-ray structure factor. This was obtained using
the following relationship
\begin{eqnarray}
 S(q)= & c_{Ge}^2 f_{Ge}(q)^ 2[S_{Ge-Ge}(q)-1] + 
 2c_{Ge} c_{O} f_{Ge}(q)f_{O}(q)[S_{Ge-O}(q)]  +  \nonumber \\
& c_{O}^2 f_{O}(q)^ 2[S_{O-O}(q)-1]
\end{eqnarray}
where $c_\alpha$ and $f_\alpha(q)$ represent  respectively the atomic fraction and
X-ray form factor of element $\alpha$.  $S_{\alpha \beta}(q)$ is the partial
structure factor which can be obtained via a Fourier transform of the partial
radial distribution functions:
\begin{eqnarray}
  S(q)_{\alpha \beta} = &
\delta_{\alpha \beta} +  \rho \int_0^{\infty} 4\pi r^2 
\frac{sin(qr)}{qr}[g_{\alpha \beta}(r)-1]dr.
\end{eqnarray}
The $q$-dependent X-ray form factors are calculated from
\begin{equation}
f_\alpha (q)=\sum_{i=1}^{4} a_{\alpha, i}\exp\;[-b_{\alpha, i} (q/4\pi)^2]+d_\alpha.
\end{equation}
The parameters we used for the calculation of these form factors factors are
reported in table \ref{table1}; note they are consistent with an ionic
representation of the distribution of electrons ({\it i.e.} Ge$^{4+}$ and
O$^{2-}$). Figure \ref{figure3} has been organised in a similar way to Fig. 1 in ref.
\cite{guthrie2004a} in order to facilitate a comparison; in particular, $S(q)$ has
been normalized by :
\begin{equation}
(\sum_{\alpha=1}^2 c_\alpha f_\alpha(q))^2.
\end{equation}
The six patterns are reported at the densities corresponding approximately to the following experimental pressures: 0, 3, 5, 7, 10, 15 GPa. In the right-hand panel of figure \ref{figure3} we report the calculated neutron diffraction patterns at the same densities; these are obtained from the same expressions but with  $f_\alpha(q)$ replaced with $b_\alpha$, the neutron diffraction length \cite{salmon2007a}. In this figure we also show some experimental data from the recent study by Drewitt and co-workers \cite{drewitt2010a} at densities close to the simulation ones. These data were collected at the following pressures: 2.2, 4.9 and 8.0 GPa.

\begin{table}
\caption{\label{table1} Parameters used for the form factors \cite{tablesrayonsx, tokonami1965a}. }
\begin{tabular}{| c | c | c | c | c | c | c | c | c | c |}
\hline
           &   a$_1$  &  b$_1$ & a$_2$  &  b$_2$ & a$_3$  &  b$_3$ & a$_4$  &  b$_4$ & d  \\
\hline
\hline
O$^{2-}$    & 4.174  & 1.938 & 3.387 & 4.145 & 1.202 & 0.228 & 0.528 & 8.285 & 0.706 \\
\hline
Ge$^{4+}$ & 4.758 &  7.831 & 3.637 & 30.05 & 0         & 0        &  0         & 0        &   1.594\\
\hline
\end{tabular}
\end{table}

By comparing figure \ref{figure3} with Fig. 1 in ref. \cite{guthrie2004a}, it can be readily appreciated that
the agreement with the experimental X-ray structure factors is excellent at all
pressures. Our data show that the First Sharp Diffraction Peak position (FSDP)
shifts toward higher wavenumbers upon pressure increase until it merges with the
principal peak, in  agreement with the data in ref. \cite{guthrie2004a}.
From the fact that our simulations are in good agreement with the  x-ray structure factors from \cite{guthrie2004a}
as well as the the more recent neutron  data  \cite{drewitt2010a} and the bond-lengths from the most recent EXAFS \cite{vaccari2009a} we conclude that the different sets of experimental data themselves are mutually consistent (see also discussion below).
The difference in the {\em interpretation} of the X-ray data by Guthrie et al. \cite{guthrie2004a} (i.e. sharp vs. smooth tetrahedral to octahedral structure) is therefore likely
to be due to the difficulty of extracting good coordination numbers from noisy data.

The comparison with the neutron data from  Guthrie {\it et al.} \cite{guthrie2004a} is inhibited by the noisiness of the experimental data but the data of Drewitt {\it et al}  \cite{drewitt2010a} is of higher quality. For densities above $\rho \sim$  5.0 g/cm$^3$, corresponding to an experimental pressure of 8-9 GPa, there are, to the best of the authors' knowledge, no available neutron data. Our simulations are consistent with the main trends observed in the neutron work, namely a diminution in the height and a shift to higher $q$ of the FSDP with increasing pressure. However, the agreement between simulations and experiment is less good than in the X-ray case. The simulations appear to overestimate the height of the FSDP, especially at low pressure, and also to underestimate the amplitude and sharpness of the principal peak at about 2.5 \AA$^{-1}$. In the total neutron structure factor of GeO$_2$ there is a near-total cancellation between the different partial structure factors in the vicinity of the principal peak, so that the ``principal" peak in the total $S(q)$ is of much lower amplitude than in the individual Ge or O partials. The discrepancy we noted above presumably reflects a small error in position or width of the principal peak in these partials. The FSDP on the other hand reflects the intermediate range order in the glass and the implication of a disagreement in this domain is that the quenching method we have used to prepare the glass does not allow this structure to develop in the same way as in the experimental material. In support of this contention, we note that the $S(q)$s calculated from the cold-compressed and densified samples differed most in this region of $q$-space.  The position of the FSDP obtained from the neutron structure factors is shown in the bottom part of figure \ref{figure3} and increases linearly with density. The corresponding values from the Drewitt {\it et al} work are also shown.
\newline

Overall then, our simulations agree well with the available pressure-dependent diffraction data when we compare directly with the experimentally measured quantity, $S(q)$. There is also good agreement with the density dependence of the nearest-neighbour separation obtained from analysis of the  EXAFS data. The examination of the local structure in the simulations supports the conclusion that  the transition from a tetrahedral to an octahedral glass is smooth and gradual and probably not completed even at pressures as high as 15 GPa.

\section{Liquid germania}

\begin{figure}
\begin{center}
\resizebox{\textwidth}{!}{\includegraphics{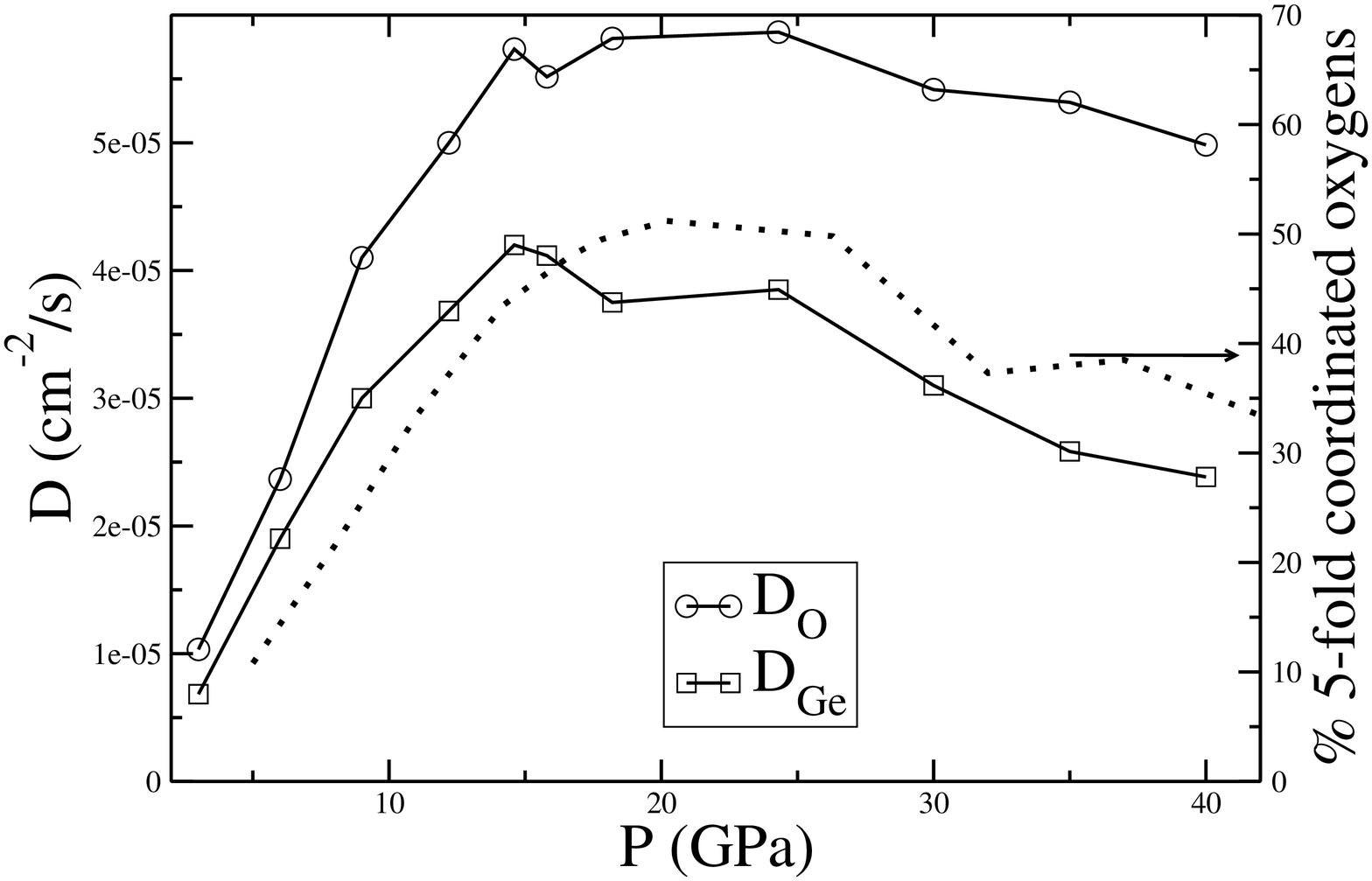}}
\end{center}
\caption{ Simulated diffusion coefficients on liquid (T = 4000 K) germania and percentage of GeO$_5$ units as a function of increasing pressure.}
\label{figure4}
\end{figure}

The high-pressure behavior of liquid germania  presents some differences compared
with the glassy case. Although a similar increase in mean coordination number is
observed, both the Ge-O coordination number and Ge-O distances (not shown) increase
more smoothly as pressure is raised. The increase in coordination number begins as soon as the pressure in increased.
In figure \ref{figure4} we show the oxygen and germanium self diffusion coefficients as a function of increasing pressure for
liquid germania at 4000 K. It can be seen that the diffusion coefficient exhibits the same anomalous
maximum in the P = 15-25 GPa range as observed in other tetrahedral-network materials
(like silica and water \cite{mcmillan2009a}). In the case of liquid silicates, it was unambiguously shown by Angell {\it et al.} that such a pressure enhancement of ion mobilities was linked to the formation of fivefold coordinated silicon ions, which act as defects in the original tetrahedral network structure.~\cite{angell1982a}. The proportion of GeO$_5$ polyhedra with each coordination number is shown on figure \ref{figure4}, and we can see that its variation with pressure matches exactly the variation of the diffusion coefficients, in agreement with the silica situation.

\section{Conclusions}
In this work we used a new polarizable interaction  potential to study the high
pressure behaviour of glassy and liquid germania.  For glassy germania, we were able to reproduce
all the experimental structural information by making comparisons at the same density. The density-evolution of the Ge-O coordination
number, bond distance, X-ray and neutron structure factors was in good agreement with
the data from refs. \cite{vaccari2009a,guthrie2004a, drewitt2010a}. The only observed shortcoming
was the inability of our simulations to properly reproduce the equation of state for glassy germania.
This is most likely due to the relatively short time-scale available to computer simulations.
From our data and from a comparison with the available experimental evidence, it can be concluded
that the transition from a tetrahedral to an octahedral glass is smooth and gradual and probably
not completed at pressures as high as 15 GPa. In view of our results, it seems likely that the differences between the mechanisms which have been proposed to interpret this data  (such as sharp {\it vs.} smooth tetrahedral to octahedral transitions) is due to the difficulty
of extracting reliable coordination number from noisy data. Finally, a study of the  percentage concentration of
GeO$_4$, GeO$_5$, GeO$_6$ units shows that as density is increased, these are present in different proportions but a
state with GeO$_5$ units only was not observed.
\newline
As for liquid germania, our results are mainly  predictive  due to the lack of
experimental data, and of interest for the comparison with the behaviour of other systems which are tetrahedrally coordinated at ambient pressure. We find that liquid germania undergoes a
transition from a fourfold- to a sixfold-coordinated phase and that this transition
is smoother than the one observed in the glassy phase. Most importantly, we find
that the diffusivity in germania behaves anomalously as a function of pressure,
showing a maximum in the P = 15-25 GPa range, as found in other tetrahedral-network
liquid such as silica and water. The cause of this behaviour is traced back to the
presence of pentahedrally-coordinated Ge ions.
\newline

\section*{Acknowledgements}
The authors wish to thank Marco Vaccari for sending us the EXAFS data on germania and Philip
Salmon for providing us with his neutron data and for fruitful discussions.
DM thanks the Moray Endowment Fund of the University of Edinburgh for the purchase of a workstation.
DM also wishes to thank the EPSRC, School of Chemistry,
University of Edinburgh, and the STFC CMPC for his PhD funding.  \newline

The work has been performed under the HPC-EUROPA2 project (project number: 228398) with
the support of the European Commission - Capacities Area - Research Infrastructures.

\section*{References}
\bibliography{references}

\end{document}